\documentclass[]{aastex}

\usepackage{emulateapj5}
\usepackage{onecolfloat}
\usepackage{graphicx}
\usepackage{fancyheadings}
\usepackage{ulem}
\usepackage{rotating}
\usepackage{lscape}

\newcommand{\cmjj}{\mbox{${\rm cm^{-2}}$}}
\newcommand{\etal}{et al.}
\newcommand{\hI}{\mbox{${\rm H\ I}$}}
\newcommand{\kms}{\mbox{km\ s${^{-1}}$}}
\newcommand{\lya}{\mbox{${\rm Ly}\alpha$}}

\newcommand{\ibid}{\underline{\makebox[0.5in]{}}.}

\begin{document}

\twocolumn[%
\lefthead{Chen \etal}
\righthead{Galaxy--\lya\ Absorbers Cross-Correlation Function}

\slugcomment{Accepted for publication in the Astrophysical Journal Letters}

\title{Probing the IGM-Galaxy Connection Toward PKS0405$-$123 II : A 
Cross-Correlation Study of \lya\ Absorbers and Galaxies at $z<0.5$\altaffilmark{1}}
\author{HSIAO-WEN CHEN\altaffilmark{2}, 
JASON X.\ PROCHASKA\altaffilmark{3}, 
BENJAMIN J.\ WEINER\altaffilmark{4}, 
JOHN S.\ MULCHAEY\altaffilmark{5}, and
GERARD M.\ WILLIGER\altaffilmark{6}}


\begin{abstract}

  We present a pilot study of the clustering properties of \lya\ absorbers with
respect to known galaxies based on 112 \lya\ absorbers and 482 galaxies 
identified at $z<0.5$ along the sightline toward PKS0405$-$123.  
The principal goal is to determine the origin of \lya\ absorbers based on their
cross-correlation amplitude with known galaxies and investigate a possible 
$N(\hI)$ dependence of the cross-correlation function.  The main results of our
study are as follows.  (1) The cross-correlation function $\xi_{ga}$ measured 
using only emission-line dominated galaxies and \lya\ absorbers of 
$\log\,N(\hI)\ge 14$ shows a comparable strength to the galaxy auto-correlation
function $\xi_{gg}$ on co-moving, projection distance scales $< 1\,h^{-1}$ Mpc,
while there remains a lack of cross-correlation signal when using only 
absorption-line dominated galaxies.  This signifies a morphology-dependent 
$\xi_{ga}$ and indicates that strong absorbers of $\log\,N(\hI)\ge 14$ and 
emission-line galaxies reside in the same halo population. (2) A 
maximum-likelihood analysis shows that \lya\ absorbers of $\log\,N(\hI)< 13.6$ 
are consistent with being more randomly distributed with respect to known 
galaxies.  Finally, (3) we find based on this single sightline that the 
amplitude of $\xi_{ga}$ does not depend sensitively on $N(\hI)$ for strong 
absorbers of $\log\,N(\hI)\ge 13.6$.

\end{abstract}

]
\altaffiltext{1}{Based on observations with the NASA/ESA Hubble Space
Telescope, obtained at the Space Telescope Science Institute, which is operated
by the Association of Universities for Research in Astronomy, Inc., under NASA
contract NAS5--26555.}

\altaffiltext{2}{Hubble Fellow at the MIT Kavli Institute for Astrophysics and 
Space Research, Cambridge, MA 02139-4307, {\tt hchen@space.mit.edu}}

\altaffiltext{3}{UCO/Lick Observatory; University of California, Santa
  Cruz, Santa Cruz, CA 95064, {\tt xavier@ucolick.org}}

\altaffiltext{4}{Department of Astronomy, University of Maryland, College Park,
MD 20742-2421, {\tt bjw@astro.umd.edu}}

\altaffiltext{5}{Observatories of the Carnegie Institution of Washington, 813 
Santa Barbara Street, Pasadena, CA 91101, U.S.A., {\tt mulchaey@ociw.edu}}

\altaffiltext{6}{Department of Physics \& Astronomy, Johns Hopkins 
University, Baltimore, MD 21218, {\tt williger@pha.jhu.edu}}


\section{INTRODUCTION}

  The forest of \lya\ absorption line systems observed in the spectra of
background QSOs offers a sensitive probe of the tenuous, large-scale baryonic 
structure that is otherwise invisible (e.g.\ Rauch 1998).  Over the last 
several years, various numerical simulations have predicted that approximately 
40\% of the total baryons at redshift $z=0$ reside in diffuse intergalactic gas
of temperature $T < 10^5$ K that gives rise to the \lya\ absorbers (e.g.\ 
Dav\'e \etal\ 2001).  Indeed, Penton, Stocke, \& Shull (2002, 2004) argue that 
\lya\ absorbers of neutral hydrogen column density $N(\hI) \le 10^{14.5}$ 
\cmjj\ may contain between $20-30$\% of the total baryons in the nearby 
universe based on observations of the local \lya\ forest and a simple 
assumption of the cloud geometry.  Furthermore, their comparison between 
galaxies and absorbers along common lines of sight implies that roughly 20\% of
low-redshift \lya\ absorbers originate in regions where no luminous galaxies 
are found within $2\ h^{-1}$ Mpc radius.  They conclude that $\approx 5$\% of 
all baryons are in voids.

  Whether or not \lya\ absorbers trace the typical galaxy population bears 
directly on the efforts to locate the missing baryons in the present-day 
universe (Persic \& Salucci 1992; Fukugita, Hogan, \& Peebles 1998), and
to apply known statistical properties of the \lya\ absorbers for constraining 
statistical properties of faint galaxies.  This issue remains, however, 
unsettled (Lanzetta \etal\ 1995; Stocke \etal\ 1995; Chen \etal\ 1998, 2001; 
Penton \etal\ 2002).  While nearly all galaxies within 180 $h^{-1}$ kpc 
physical radius of the QSO lines of sight have associated \lya\ absorbers of 
$N(\hI) \ge 10^{14}$ \cmjj\ that are less than a velocity separation $v = 250$ 
\kms\ away along the sightline (Chen \etal\ 1998, 2001), not all \lya\ 
absorbers have a galaxy found within 1 $h^{-1}$ Mpc physical distance to a 
luminosity limit of $0.5\, L_*$ (Morris \etal\ 1993; Tripp \etal\ 1998).  
Despite the discordant interpretations in associating \lya\ absorbers with 
galaxies, the \lya\ absorbers (particularly those of $N(\hI) \ge 10^{14}$ 
\cmjj) are consistently found to be correlated with known galaxies (Morris 
\etal\ 1993; Bowen, Pettini, \& Blades 2002) but at a weaker clustering 
strength than what is measured between field galaxies.  The weak 
cross-correlation amplitude between galaxies and absorbers has led to the 
conclusion that the \lya\ absorbers trace large-scale filamentary structures, 
rather than individual galactic halos (e.g.\ Grogin \& Geller 1998; Tripp 
\etal\ 1998).

\begin{figure*}[ht]
\begin{center}
\includegraphics[scale=0.4]{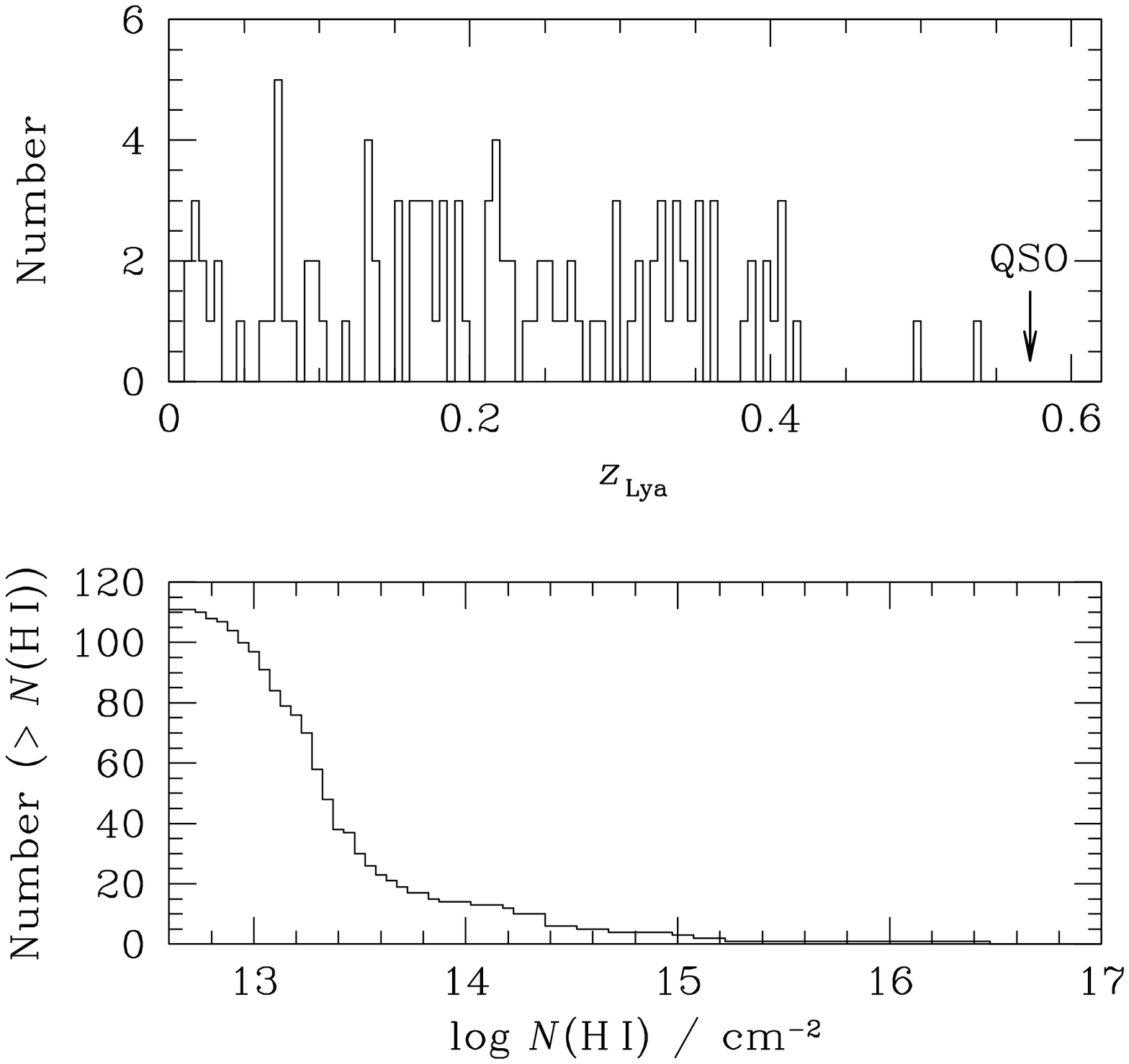}
\includegraphics[scale=0.4]{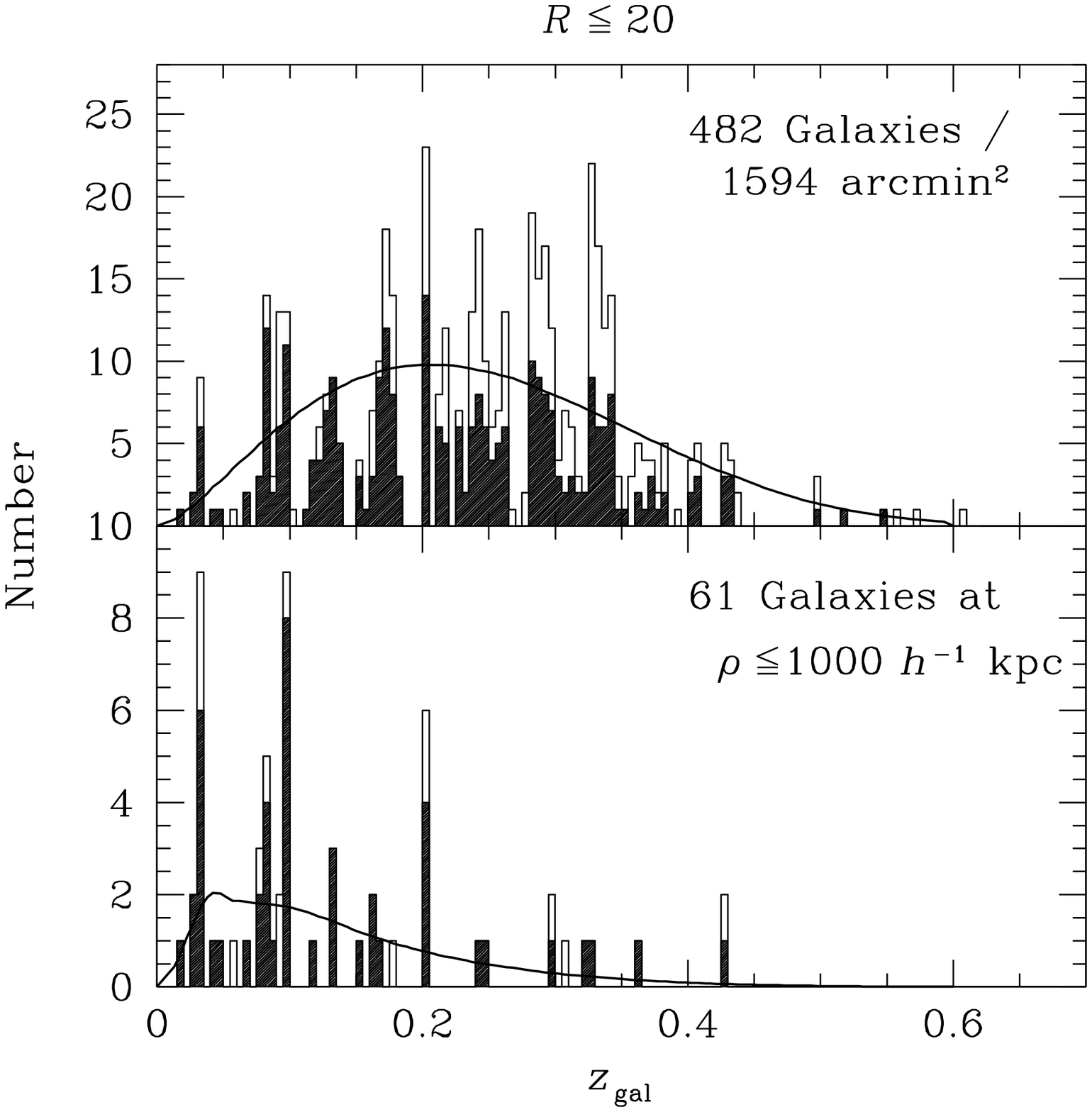}
\caption{Left: The top panel shows the redshift distribution of all 112 \lya\ 
absorbers identified along the sightline toward PKS0405$-$123.  The detection
limit of \lya\ absorption lines is uniform at $N(\hI)=13.3$ across the entire 
redshift range over $z=0.002-0.423$ for absorbers with Doppler parameter $b\le
40$ \kms, and better than $N(\hI)=13.1$ over $z=0.02-0.243$.  \lya\ absorbers 
at $z>0.423$ are found in the FOS spectra retrieved from the HST data archive.
The bottom panel shows the cumulative $N(\hI)$ distribution of the absorbers.  
Right: Redshift distribution of galaxies with $R\le 20$ in the field around 
PKS0405$-$123.  The top panel shows all galaxies over the 1600 arcmin$^2$ sky 
area centered at roughly at the background QSO, for which we have obtained 
robust redshift measurements.  The bottom panel shows galaxies within 1 
$h^{-1}$ Mpc comoving radius in each redshift bin.  In both panels, the open 
histogram represents galaxies of all spectral types, the shaded histogram 
represents emission-line dominated galaxies, and the solid curve shows the 
selection function of our redshift survey, which is 90\% complete at $R=20$.}
\end{center}
\end{figure*}

  However, galaxies are known to be a biased tracer of the underlying dark 
matter halos and the bias depends on morphology and luminosity (see e.g.\ Davis
\& Geller 1976; Loveday \etal\ 1995; Madgwick \etal\ 2003).  The weaker 
cross-correlation amplitude observed between galaxies and \lya\ absorbers, when
compared with galaxy auto-correlation studies, can possibly be explained by a 
difference in galaxy bias.  In this {\em Letter}, we present a pilot study of 
the galaxy--\lya\ absorber cross-correlation function based on the \lya\ 
absorbers and galaxies identified along the sightline toward PKS0405$-$123.  
The primary objectives of the study are (1) to compare the galaxy--absorber 
cross-correlation function with galaxy auto-correlation function for galaxies 
of different properties, and (2) to obtain a quantitative estimate of the 
$N(\hI)$ threshold below which the \lya\ absorbers no longer trace the 
large-scale galaxy structures.

\section{THE LY$\alpha$-ABSORBER AND GALAXY SAMPLES}

\renewcommand{\thefootnote}{\fnsymbol{footnote}}
\setcounter{footnote}{1}

  Exquisite echelle spectra of the QSO PKS0405$-$123 ($z_{\rm qso}=0.5726$) 
have been obtained by the STIS GTO team (${\rm PID}=7576$), using the Space 
Telescope Imaging Spectrograph (STIS) on board the Hubble Space Telescope (HST)
with a $0.2'' \times 0.06''$ slit and the E140M grating ($R = 45,800$ or 6.7 
\kms) for a total exposure time of 27,208 s.  The final stacked spectrum covers
a spectral range spanning from 1140 to 1730 \AA\ and has on average $S/N 
\approx 7$ per resolution element over the entire spectral region.  Additional 
spectra obtained using the Faint Object Spectrograph (FOS) have been included 
for finding strong absorption lines $\log\,N(\hI) > 14$ outside of the 
STIS/E140M spectral range at $0.423 \le z_{\rm Ly\alpha}\le 0.557$.  Detailed 
data reduction and absorption line measurements are presented in Williger 
\etal\ (2005).  We have identified 112 \lya\ absorption lines with 
$\log\,N(\hI) = 12.5 - 16.5$ at greater than 4-$\sigma$ significance level over
$0.01 < z_{\rm Ly\alpha} < 0.54$.  The redshift distribution and cumulative 
$N(\hI)$ distribution functions of the \lya\ absorber sample along this 
sightline are presented in the left panels of Figure 1.

  We have carried out a redshift survey of galaxies in the field around 
PKS0405$-$123, using the WFCCD multi-slit spectrograph mounted on the 2.5 m du 
Pont telescope at the Las Campanas Observatory.  The primary goal of the 
project is to study the large-scale correlation between galaxies and QSO 
absorption-line systems at low redshift.  We therefore aim to obtain a 
complete, magnitude-limited survey of galaxies over a field size that is 
comparable to the characteristic correlation length of field galaxies at $z\sim
0.2$.  Detailed descriptions of the galaxy survey and reduction of the 
spectroscopic data are presented in Prochaska \etal\ (2005, in preparation).
In summary, we have observed 535 galaxies of $R$-band magnitude $R\le 20$ over 
a 1600-arcmin$^2$ non-contiguous region centered on the background QSO.  
Robust redshift measurements are available for 482 of these galaxies.  Our 
survey is sensitive to galaxies of 0.4 $L_*$ at $z=0.2$ within 5 $h^{-1}$ Mpc 
comoving radius from the QSO lines of sight.\footnote{Throughout the paper, we
adopt a $\Lambda$ cosmology, $\Omega_{\rm M}=0.3$ and $\Omega_\Lambda = 0.7$, 
with a dimensionless Hubble constant 
$h = H_0/(100\ {\rm km} \ {\rm s}^{-1}\ {\rm Mpc}^{-1})$.}

  The redshift distribution of all 482 galaxies identified in the field around 
PKS0405$-$123 is presented in the top-right panel of Figure 1 .  Redshifts are 
determined using multiple spectral line features identified in the data. 
Typical redshift measurement errors are $\Delta z = \pm 0.0002$ or $\sim 40$ 
\kms.  We also perform a $\chi^2$ analysis, which compares the galaxy spectra 
with a model established from a linear combination of four principal components
from the Sloan Digital Sky Survey galaxy sample (Schlegel \etal\ 2005, in 
preparation).  This procedure allows us to characterize each galaxy based on 
its emission- and absorption-line properties.  We found that 284 of all 
galaxies in our survey have emission-line dominated spectral features.  The 
redshift distribution of 61 galaxies found within a comoving radius of $\rho = 
1\ h^{-1}$ Mpc from the QSO line of sight is presented in the bottom-right 
panel of Figure 1 .  The shaded histogram represents emission-line galaxies, 
while the open histogram represents all galaxies identified in the same volume.
The solid curve in each panel shows the selection function of our survey.

\section{ANALYSIS}

\subsection{The Galaxy-Absorber Cross-correlation Function}

  In the formalism developed by Mo \& White (1996), the halo-mass 
cross-correlation function $\xi_{hm}(r)$ is related to the mass 
auto-correlation function $\xi_{mm}(r)$ by a mass-dependent bias $b(M)$, as 
$\xi_{hm}(r) = b(M)\,\xi_{mm}(r)$.  Comparing the amplitudes of the 
galaxy-absorber cross-correlation function $\xi_{ga}$ and the galaxy 
auto-correlation function $\xi_{gg}$ therefore yields an estimate of the 
absorber halo mass, if the mean halo mass of the galaxies is known.  This 
approach, which has been applied to study the origin of intermediate-redshift 
Mg\,II absorbers (Bouch\'e, Murphy, \& P\'eroux 2004), has an important 
advantage in that the complicated selection function of a galaxy survey impacts
both $\xi_{ga}$ and $\xi_{gg}$ in the same way and is therefore cancelled out 
in the calculation.

  In principle, the correlation function is multi-variate.  It depends on the 
velocity separation between galaxy-absorber or galaxy-galaxy pairs $v$ along 
the line of sight, the comoving projected distance of a pair $\rho$, and the 
underlying halo mass of the source, for which we use the spectral properties of
the galaxies as a proxy.  In agreement with previous findings of Lanzetta, 
Webb, \& Barcons (1997), who considered $\xi_{ga}$ as a function of $v$ and 
$\rho$, we find that $\xi_{ga}(v)$ measured from the sample of 112 \lya\ 
absorbers and 482 galaxies identified along the sightline toward PKS0405$-$123 
weakens toward larger $\rho$ and shows no signal beyond $\rho=1\,h^{-1}$ Mpc. 
Here we focus on a 1-D correlation function measured versus $v$, and consider 
only pairs at $\rho \le\, 1\,h^{-1}$ Mpc.  We compare $\xi_{ga}(v)$ measured 
for \lya\ absorbers of different $N(\hI)$ with $\xi_{gg}(v)$ measured 
separately for all galaxies, absorption-line dominated galaxies, and 
emission-line dominated ones (Figure 2).  Error bars in each velocity 
separation bin are estimated assuming Poisson statistics.

\begin{figure*}
\begin{center}
\includegraphics[scale=0.65]{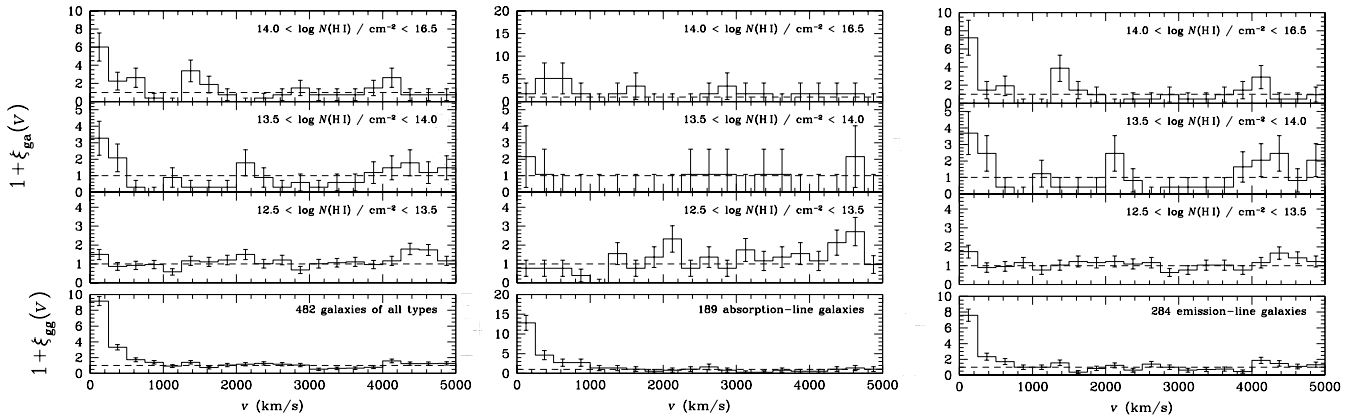}
\caption{Galaxy-\lya\ absorber cross-correlation function $\xi_{ga}(v)$ versus
velocity separation $v$ of each pair along the line of sight in the top three 
panels for \lya\ absorbers of different $N(\hI)$, and galaxy-galaxy
auto-correlation function $\xi_{gg}(v)$ versus $v$ in the bottom panel.  We 
note that the limits on the $y$-axis for weak absorbers has been rescaled for a
better visibility of the cross-correlation signal at $v<500$ \kms.  For 
$\xi_{ga}$, we consider all 61 galaxies found within $\rho = 1\ h^{-1}$ Mpc 
co-moving radius to the sightline in the left panels, and 15 absorption-line 
dominated and 46 emission-line galaxies at $\rho < 1\ h^{-1}$ Mpc in the middle
and right panels, respectively.  For $\xi_{gg}$, we consider all pairs of 
different spectral type with a co-moving impact separation $< 1\ h^{-1}$ Mpc.}
\end{center}
\end{figure*}

\subsection{Clustering of \lya\ Absorbers with Galaxies of Different Spectral 
Morphology}

  Three distinct features are qualitatively apparent in the left panels of 
Figure 2.  First, a strong cross-correlation signal in $\xi_{ga}(v)$ for \lya\ 
absorbers of $\log\,N(\hI) > 13.5$ is present over velocity separation $v < 
250$ \kms.  The signal has $>4-\sigma$ level of significance when
compared with random galaxy and absorber pairs found at large velocity 
separation along the sightline, confirming that these absorbers are not 
randomly distributed with respect to galaxies.  Second, this cross-correlation 
signal in $\xi_{ga}(v)$ becomes weaker for absorbers of lower $N(\hI)$, 
suggesting a correlation between $N(\hI)$ and the mass scale of the underlying 
dark matter halos traced by these absorbers.  In addition, the absence of a 
cross-correlation signal for \lya\ absorbers of $\log\,N(\hI) < 13.5$ indicates
that these weak absorbers do not trace the spatial distribution of galaxies.  
Third, we find that when considering all galaxies together the amplitude in 
$\xi_{gg}$ is more than 50\% stronger than the amplitude in $\xi_{ga}$, even 
for the strong absorbers in the top panel.  This is consistent with previous 
findings, which have been interpreted as \lya\ absorbers tracing large-scale 
filamentary structures rather than individual galactic halos (e.g.\ Grogin \& 
Geller 1998; Tripp \etal\ 1998).

  To examine the nature of \lya\ absorbers, we repeat the analysis including 
galaxies of different spectral properties.  We divide the galaxies into two 
subsamples depending on whether their spectra are dominated by absorption or 
emission features.  Our choice is guided by the fact that absorption-line 
dominated galaxies are more strongly clustered than emission-line dominated 
galaxies (e.g.\ Madgwick \etal\ 2003).  The results of the correlation analysis
including only absorption-line and only emission-line dominated galaxies are 
presented respectively in the middle and right panels of Figure 2.  We find 
that $\xi_{ga}(v)$ measured using only absorption-line dominated galaxies is 
consistent with zero on all velocity separation scales for the entire \lya\ 
absorber sample, contrary to the enhanced $\xi_{gg}$ measured for 
absorption-line dominated galaxies only.  In addition, we find that 
$\xi_{ga}(v)=6.2\pm 1.9$ for \lya\ absorbers of $\log\,N(\hI) \ge 14$ and 
emission-line galaxies are consistent with $\xi_{gg}(v)=6.6\pm 0.8$ for 
emission-line galaxies to within 1-sigma uncertainties, indicating that these 
strong \lya\ absorbers and emission-line dominated galaxies found in our 
redshift survey trace the same halo population.  The weaker $\xi_{ga}(v)$ at 
lower $N(\hI)$ observed in the right panels also suggests that weaker absorbers
may originate in still lower mass halos that are loosely clustered with these 
known galaxies.

\subsection{$N(\hI)$-Dependent Clustering Strength of \lya\ Absorbers with 
Galaxies}

  Much of the past dispute regarding the nature of \lya\ absorbers originates
in the different $N(\hI)$ regimes considered by different authors.  It appears
that while most strong absorbers of $N(\hI)\ge 14$ can be explained by the 
known galaxy population (e.g.\ Chen \etal\ 1998, 2001), the majority of weaker 
absorbers remain as a poor tracer of galaxies (e.g.\ Penton \etal\ 2002, 2004).
This is also observed in the declining cross-correlation amplitude with 
$N(\hI)$ as presented in Figure 2.  We perform a maximum likelihood 
analysis to obtain a quantitative estimate of the $N(\hI)$ threshold below 
which the \lya\ absorbers no longer trace known galaxies.

  We first formulate the cross-correlation function as
\begin{equation}
\xi_{ga}(v)=A\,\times\,\left(\frac{N({\rm H\,I})}{N_0}-1\right)^\beta\,\times\,\exp\left[-\frac{1}{2}\left(\frac{v}{v_0}\right)^2\right]
\end{equation}
if $N(\hI)\ge N_0$ and $\xi_{ga}(v)=0$ otherwise.  We model the velocity 
dispersion function of the galaxy and absorber pairs using a gaussian of FWHM 
$v_0$.  According to the model described in Equation (1), we expect that the 
cross-correlation amplitude is scaled with $N(\hI)$ for absorbers of $N(\hI)\ge
N_0$ following a power-law function, and vanishes for weaker absorbers.  For 
each \lya\ absorber $i$ of $N^i(\hI)$ at $z_{\rm Ly\alpha}^i$, the probability 
of finding a galaxy $j$ within the volume defined by $z_{\rm Ly\alpha}^i\pm 
\Delta z$ and $\rho_j \le 1\,h^{-1}$ Mpc is then written as
\begin{equation}
{\scriptstyle P_i(z_{\rm gal}^j) = \frac{\left\{1 + A\,\times\,\left(\frac{N^i({\rm H\,I})}{N_0}-1\right)^\beta\,\times\,\exp\left[-\frac{1}{2}\left(\frac{v_{ij}}{v_0}\right)^2\right]\right\}\times \Delta z}
{\int {\cal S}(z)\times {\left\{1+ A\,\times\,\left(\frac{N^i({\rm H\,I})}{N_0}-1\right)^\beta\,\times\,\exp\left[-\frac{1}{2}\left(\frac{v(z_{\rm gal}-z_{\rm Ly\alpha})}{v_0}\right)^2\right]\right\}}dz}},
\end{equation}
where $v_{ij}/c=[(1+z_{\rm gal})^2-(1+z_{\rm Ly\alpha})^2]/[(1+z_{\rm 
gal})^2+(1+z_{\rm Ly\alpha})^2]$ and ${\cal S}(z)$ is the redshift selection 
function of our galaxy survey within the $1\,h^{-1}$ Mpc comoving radius as 
presented in the bottom-right panel of Figure 1.  Considering all the galaxy 
and \lya\ absorber pairs as a whole, we can formulate the likelihood function 
of observing $m$ galaxies at $\rho\le 1\,h^{-1}$ Mpc from $n$ \lya\ absorbers 
of $12.5\le N(\hI)< 17$ as the following,
\begin{equation}
{\cal L}(A,N_0,\beta,v_0)=\prod_{i=1}^{n}\prod_{j=1}^{m}P_i(z_{\rm gal}^j).
\end{equation}
Finally, maximizing ${\cal L}$ with respect to $A$, $N_0$, $\beta$, and $v_0$ 
leads to best-fit values of $A=11\pm 3$, $\log\,N_0=13.6\pm0.1$, $\beta=0.0\pm 
0.1$, and $v_0=102\pm 9$ \kms, where the error of each parameter represents 
the 68\% 1-$\sigma$, 1-parameter uncertainty.  

  Adopting a power-law model $(N(\hI)/N_0)^\beta$ with $\log\,N_0=13.6$ to 
replace the step function in Equation (1) yields a best-fit $\beta=0.20 \pm 
0.02$.  The non-zero value reflects a significant difference in $\xi_{ga}$ 
between strong and weak absorbers.  We note, however, that the best-fit 
power-law model substantially underestimates $\xi_{ga}$ at $\log\,N(\hI)\ge 
14$, when extrapolating from $\log\,N(\hI)=12.5-13$, and is therefore not 
likely to represent the intrinsic shape of $\xi_{ga}$.  Adopting a 
gamma-function form following \\ $\exp(-N_0/N(\hI))\times\,(N_0/N(\hI))^\beta$ 
yields $\log N_0=13.5\pm 0.1$ and $\beta=0.0 \pm 0.1$.

\section{DISCUSSION AND CONCLUSIONS}
 
  Our pilot study of the galaxy-\lya\ absorber cross-correlation function at 
$z<0.5$ using galaxies and absorbers identified along the sightline toward 
PKS0405$-$123 has offered important insights for understanding the nature of 
\lya\ absorbers over a wide range of $N(\hI)$:

  First, the cross-correlation analysis not only confirms that \lya\ absorbers 
(particularly those of $\log\,N(\hI) \ge 14$) are not randomly distributed with
respect to galaxies, but also shows that emission-line dominated galaxies and 
these strong absorbers have comparable clustering strength.  The difference in 
the auto-correlation amplitudes observed for absorption- and emission-line 
dominated galaxies confirms that absorption-line galaxies trace higher-mass 
halos with stronger clustering strength than emission-line galaxies.  We 
therefore conclude that \lya\ absorbers of $\log\,N(\hI)\ge 14$ and 
emission-line dominated galaxies found in our redshift survey do indeed trace 
the same halo population.  A quantitative estimate of the halo mass scale 
traced by the \lya\ absorbers will require a detailed clustering analysis of 
these emission-line galaxies over a large volume.

  Second, we observe a likely transition in the clustering strength of \lya\ 
absorbers with respect to galaxies at $\log\,N(\hI) = 13.6$, above which the
amplitude of $\xi_{ga}(v)$ is comparable to the emission-line galaxy 
auto-correlation function and below which $\xi_{ga}(v)$ quickly declines to 
within 2.5 $\sigma$ of zero.  The diminishing cross-correlation signal between
these weak absorbers and known galaxies suggests that they are a poor tracer of
galactic structures.  A detailed comparison with numerical simulations will 
help us to constrain the amount of baryons contained in these low-$N(\hI)$ 
\lya\ absorbers.  In addition, we do not detect a strong dependence of 
$\xi_{ga}$ on $N(\hI)$ at $\log\,N(\hI)\ge 13.6$.  It would be interesting to 
see whether $\xi_{ga}$ remains insensitive to $N(\hI)$ with a larger galaxy and
absorber pair sample over multiple lines of sight.

  With a larger sample of galaxies and absorbers over multiple lines of sight, 
we hope to obtain a better constraint for the galaxy-\lya\ absorber 
cross-correlation function in the near future.  In addition, we will extend 
this analysis to O\,VI absorbers identified along these sightlines, in order to
locate the warm-hot intergalactic medium probed by these absorbers (e.g.\ 
Mulchaey \etal\ 1996; Tripp, Savage, \& Jenkins 2001).

\acknowledgments
 
  We appreciate the expert assistance from the staff of the Las Campanas 
Observatory.  It is a pleasure to thank Steve Shectman and Rob Simcoe for 
helpful discussions.  We thank the referee, Joop Schaye, for a swift review
and helpful comments.  H.-W.C. acknowledges support by NASA through a Hubble 
Fellowship grant HF-01147.01A from the Space Telescope Science Institute, which
is operated by the Association of Universities for Research in Astronomy, 
Incorporated, under NASA contract NAS5-26555.  JXP acknowledges support from 
NASA grants NAG5-12496 and NAG5-12743 through Carnegie Observatories and 
UCO/Lick Observatory.


\end{document}